\documentclass{article}
\usepackage{spconf,amsmath,graphicx,hyperref}
\usepackage{lipsum} 
\usepackage{booktabs}
\usepackage[dvipsnames]{xcolor}
\usepackage{cite}
\usepackage{enumitem}

\title{Controllable Audio-Visual Viewpoint Generation \\ from 360$^{\circ}$ Spatial Information}
\name{Christian Marinoni$^{*}$
\thanks{
$^{*}$ Equal contribution.}, Riccardo F. Gramaccioni$^{*}$, Eleonora Grassucci$^{}$, and Danilo Comminiello$^{}$}
\address{$^{}$Sapienza University of Rome, Italy}
\begin{document}
\ninept
\maketitle
%
\begin{abstract}
The generation of sounding videos has seen significant advancements with the advent of diffusion models. However, existing methods often lack the fine-grained control needed to generate viewpoint-specific content from larger, immersive 360-degree environments. This limitation restricts the creation of audio-visual experiences that are aware of off-camera events. To the best of our knowledge, this is the first work to introduce a framework for controllable audio-visual generation, addressing this unexplored gap. Specifically, we propose a diffusion model by introducing a set of powerful conditioning signals derived from the full 360-degree space: a panoramic saliency map to identify regions of interest, a bounding-box-aware signed distance map to define the target viewpoint, and a descriptive caption of the entire scene. By integrating these controls, our model generates spatially-aware viewpoint videos and audios that are coherently influenced by the broader, unseen environmental context, introducing a strong  controllability that is essential for realistic and immersive audio-visual generation.
We show audiovisual examples proving the effectiveness of our framework
.
\end{abstract}
\begin{keywords}
Audio-visual generation, 360-degree video, diffusion models, controllability, spatial awareness.
\end{keywords}
%
\section{Introduction}
\label{sec:intro}
\begin{figure}[!t]
    \centering
    \includegraphics[width=\linewidth]{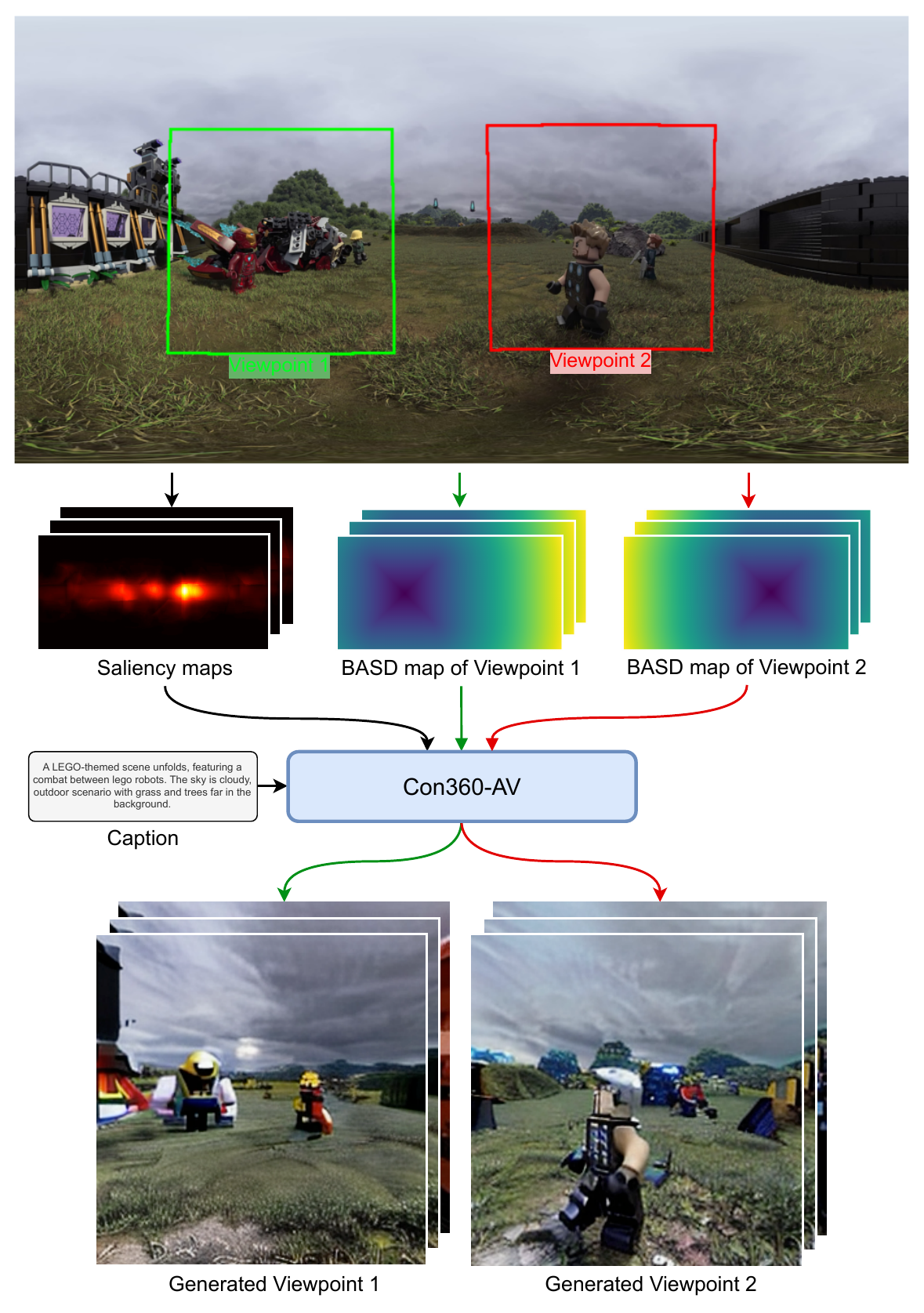}
    \caption{The Con360-AV generation process. The model generates a target viewpoint conditioned on three inputs: global 360° saliency maps, a scene-wide text caption, and viewpoint-specific BASD maps. The figure shows how two different outputs, Viewpoint 1 (\textcolor{OliveGreen}{\textit{green}}) and Viewpoint 2 (\textcolor{Mahogany}{\textit{red}}), are generated from the same scene by providing their corresponding BASD maps.}
    \label{fig:overview}
\end{figure}
\begin{figure*}[!t]
    \centering
    \includegraphics[width=0.95\linewidth]{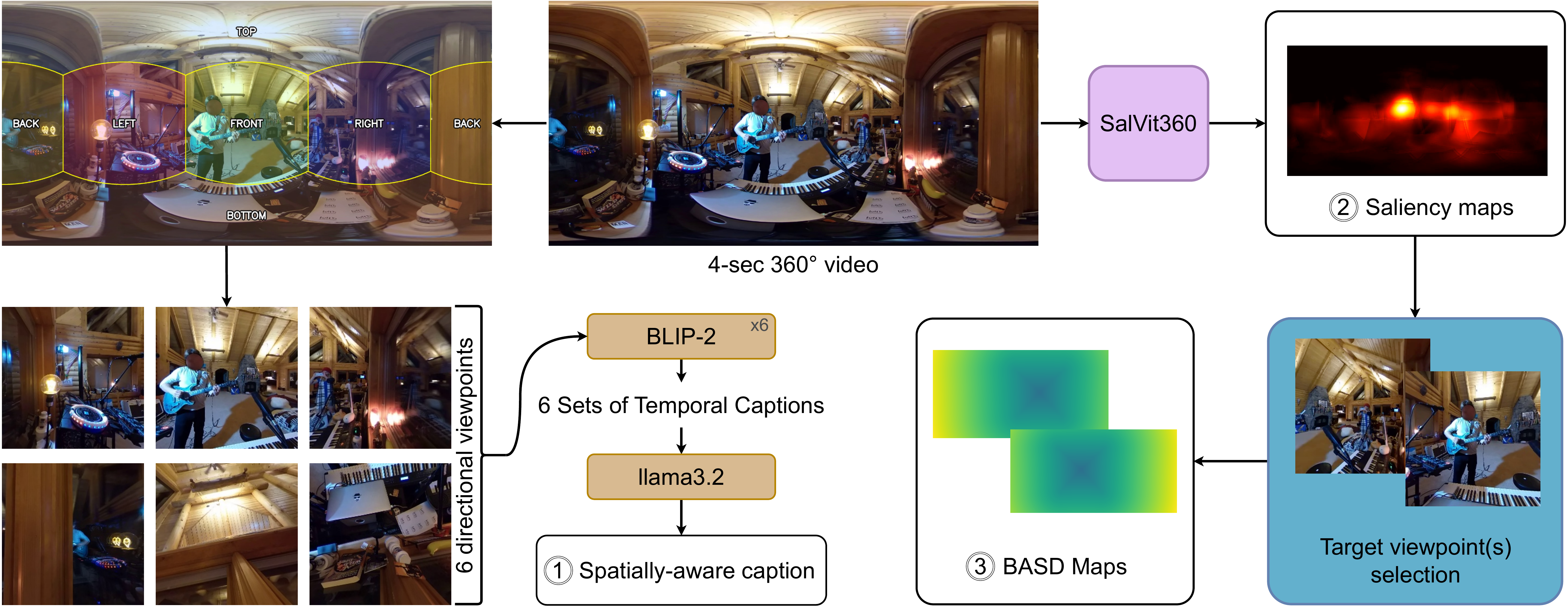}
    \caption{   
    \textbf{Derivation of the three contextual conditionings from a 360° video.} The pipeline generates a textual prompt and two visual prompts through parallel branches. \textbf{(1) Spatially-aware Caption}: For each of the six directional viewpoints (front, back, left, right, top, bottom), BLIP-2 generates a series of captions at different timestamps to capture the evolution of the scene. Llama-3.2 then synthesizes this full set of descriptions to reconstruct a single, dynamic caption summarizing the action across the entire 360° view over time. \textbf{(2) Saliency Maps}: In parallel, the full 360° video is processed by SalViT360 to generate a sequence of saliency maps that highlights the most visually important regions over time. \textbf{(3) BASD Maps}: Centroids from the saliency maps are used to define the target viewpoint and generate a corresponding BASD (Boundary-Aware Saliency Detection) map for structural guidance.
    }
    \label{fig:spatialrep}
\end{figure*}
Generative models, particularly diffusion-based approaches, have demonstrated remarkable success in synthesizing high-fidelity content across single modalities like images, video, and audio \cite{rombach2022high,guo2023animatediff,liu2023audioldm}. Many research works focused on generating one modality using another one as conditioning, as in the case of video to audio generation \cite{comunita2024syncfusion, gramaccioni2024folai, GramaccioniIJCNN2025, MarinoniIJCNN2025}.
 A significant frontier in this field is the joint generation of multiple modalities, such as synchronized audio and video (AV) \cite{liu2024syncflow, liu2025javisdit}.
When dealing with this task, spatial, semantic and temporal alignment is of fundamental importance. In \cite{ishii2024simple}, temporal synchrony is maintained mainly through two key mechanisms: timestep adjustment, which aligns the different noise schedules of audio and video to ensure the denoising processes remain mutually informative, and Cross-Modal Conditioning as Positional Encoding (CMC-PE), which provides a strong inductive bias for temporal synchronicity by directly adding time-aligned features from one modality to another. Regarding semantic information, models usually leverage state-of-the-art semantic multimodal encoders \cite{laionclap2023, cicchetti2024gramian}.
Still, none of these works guarantee spatial concurrence between audio and video. Several AV methods use sequential approaches, like MM-Diffusion \cite{ruan2023mm} and AV-DiT \cite{kim2024versatile}. 
While these methods are promising, the sequential nature of such works does not provide sufficient modularity, which is crucial for injecting diverse and complex conditioning signals.
Notably, these models operate on standard, forward-facing videos. A significant challenge emerges when dealing with immersive 360° content for generating a video of a specific field of view (FoV) while ensuring both audio and visuals remain consistent with events happening outside that view. For instance, if a band plays in a circle, the generated video of the guitarist should feature audio that correctly spatializes the sound of the drummer who is currently off-screen. Existing models lack this explicit spatial controllability.


Our approach, Con360-AV (Controllable 360° context Audio-Visual generation), addresses this limitation by unifying three key concepts from recent, yet distinct, lines of research into a coherent and novel framework.
First, for immersive media, leveraging the full 360° environmental context is essential  \cite{wang2024360dvd}. Work such as OmniAudio \cite{mao2025omniaudio} has demonstrated that conditioning on a full panorama, rather than a limited field-of-view, substantially improves the quality and spatial accuracy of generated audio by accounting for off-screen sources.
Second, saliency maps have proven effective for semantic and attentional guidance. GazeFusion \cite{zhang2024gazefusion}, for example, established that a 2D saliency map can be used to steer a model's focus, making it a valuable signal for highlighting important regions within a scene.
Finally, for precise object placement, map-based geometric representations offer robust control. For instance, 2D bounding boxes can be converted into a Bounding Box-Aware Signed Distance (BASD) map, providing a dense signal to guide object layout \cite{simoni2025bounding}.

These approaches have been developed in recent research works and not yet been combined in a unified framework for 360°-aware audio-visual generation.
To bridge this gap, we integrate these three concepts
into a novel framework for controllable and temporally aligned- sounding video generation in 360° environments. 
Our work is the first to leverage advanced spherical control concepts to generate viewpoint-specific content that is coherent with a larger, explicitly defined off-screen world.
Comprehensive spatial awareness is then guaranteed by a 360° visual saliency map, a BASD map and a global text caption describing the entire 360° scene. Such controls provide meaningful conditioning to our joint AV diffusion model. 
The experiments conducted on a selection of videos from \textit{Sphere360}, a large-scale dataset, demonstrate the effectiveness of the proposed method. 
To the best of our knowledge, this is the first research attempt that focuses on creating sounding videos that are consistently influenced by the 360° spatial context for the design of specific camera viewpoints.

This method, whose schematic functioning diagram is shown in Fig.~\ref{fig:overview} may pave the way for multiple applications in the generation of audio-visual media, such as enriching immersive VR experiences with off-camera sounds, simulating realistic ambient audio for 360° videos, or enabling interactive storytelling where the perceived soundscape adapts to the user viewpoint.
Furthermore, by conditioning the generation on the complete 360° context, our method gains explicit control over events at the viewpoint boundary, such as character entries and exits. This approach also guarantees spatio-temporal coherence, ensuring that actions remain consistent across dynamic camera movements.
%
%
\vspace{-0.235cm}
\section{Proposed Method}
\label{sec:method}
Our proposed method, Con360-AV, leverages on and extends the strong parallel-model baseline of Ishii \textit{et al.} \cite{ishii2024simple} to enable controllable audio-visual generation from a comprehensive 360° context. Exploiting two parallel, pre-trained U-Nets with CMC-PE for temporal synchronization, we introduce a conditioning module that processes and injects a rich set of spatial and semantic signals derived from the full panoramic scene. This module guides the generation of a specific field-of-view (FoV) video and its corresponding audio, ensuring they are coherent with the broader, off-screen environment.

\subsection{360° Context Representation}
To provide the model with a comprehensive understanding of the scene, we generate three distinct conditioning signals from the source 360° video.

\begin{enumerate}[label=(\arabic*),leftmargin=1.7em]
    \item \textbf{360° Saliency Maps} identify regions of interest within the panoramic scene. We first generate visual saliency maps for the 360° video, which is provided in equirectangular projection (ERP). We use a pre-trained saliency prediction model, SalViT360 \cite{rai2021salvit360}, to produce a heatmap that highlights visually significant objects and actions across the entire sphere. This signal informs the model about important events, regardless of whether they are on-screen or off-screen in the final generated viewpoint.
    \item \textbf{Bounding Box-Aware Signed Distance (BASD)} maps precisely define the target viewpoint for generation, introducing a strong geometric conditioning signal. To determine the center of our target Field of View (FoV), we first identify the centroids of the most prominent regions within the 360° visual saliency maps. This is achieved by applying normalization and a subsequent thresholding operation to highlight the main regions of interest. These centroids and their related bounding boxes correspond to the relevant subjects in the scene and, consequently, to the various viewpoints of interest.
    Each bounding box for a target FoV is then projected onto the Equirectangular Projection (ERP) frame, accounting for the inherent deformations. From this projected bounding box, we compute a 2D BASD map, where the value of each pixel represents its signed Euclidean distance to the nearest boundary of the FoV bounding box (positive inside, negative outside). This continuous map provides a powerful geometric prior, explicitly informing the model about the precise location and boundaries of the content it needs to generate.
    \item \textbf{360° Scene Caption} provides high-level semantic context. We generate a detailed description of the entire 360° scene. This is achieved through a multi-step process: first, the ERP video frames are projected onto the six faces of a cubemap. Then each of the six perspective views is fed into a pre-trained image captioning model (BLIP-2 \cite{li2023blip}) to generate six spatially-localized captions per frame. Finally, these localized descriptions are aggregated over a time interval using the Llama 3.2 Large Language Model (LLM) \cite{grattafiori2024llama}, to produce a single, coherent prompt that describes the complete 360° environment and its dynamics.
\end{enumerate}

\noindent The entire process is schematized in Fig.~\ref{fig:spatialrep}.

\subsection{Controllable Generation Module}
The three context signals are integrated into the frozen audio U-Net and video U-Net of the model via a newly introduced, trainable control module.

The saliency and BASD maps are first min-max normalized and stacked along the channel dimension. This combined map tensor is then processed by a dedicated Map Encoder, which consists of a sequence of 2D convolutional layers followed by an LSTM block to capture spatio-temporal features. The output features are downsampled using adaptive average pooling to match the spatial dimensions of the U-Net intermediate feature maps at various depths.
The features from our Map Encoder are then injected into the video and audio U-Nets using Feature-wise Linear Modulation (FiLM) \cite{perez2018film} layers. These lightweight layers apply an affine transformation to the U-Net features, modulating the generation process. Critically, these FiLM layers are inserted between the cross-modal self-attention block (from CMC-PE) and the subsequent spatial self-attention block, allowing our control signals to directly influence the spatially-aware feature representation after cross-modal synchronization has occurred. During training, we optimize only the new control modules (Map Encoder and FiLM layers) together with the original CMC-PE blocks, thereby preserving the powerful generative priors of the pre-trained audio and video sub-models.

\subsection{Audio-visual Synthesis Model}
The audio-visual synthesis model is built upon a dual U-Net architecture~\cite{ishii2024simple}, leveraging two pretrained diffusion models: AudioLDM~\cite{liu2023audioldm} for the audio stream and AnimateDiff~\cite{guo2023animatediff} for the video stream. These powerful unimodal generators are integrated into a cohesive, joint generation framework through a set of lightweight, trainable connector modules. These connectors bridge the audio and video UNets, creating a bidirectional information flow that is crucial for synchronizing the two modalities during the shared denoising process.
To achieve strong temporal alignment between the audio and video we use the CMC-PE mechanism, which is implemented through an additive conditioning scheme where features from one modality are directly injected into the other. This process occurs at the two deepest resolution levels of the U-Net architecture to capture both high-level semantic and fine-grained temporal correlations. Specifically, the trainable connector modules are inserted after the third and fourth down-sampling blocks in the encoder path, and symmetrically, after the first and second up-sampling blocks in the decoder path of each U-Net. This is also the stage where our spatial conditioning signals are integrated, allowing them to modulate the spatially-aware features after the initial cross-modal synchronization has occurred.
%
\section{Experiments}
\label{sec:experiments}
\subsection{Dataset}

For our experiments we use \textit{Sphere360} \cite{mao2025omniaudio}, a large-scale collection of 360° videos paired with the corresponding audio. To the best of our knowledge, this is the first attempt to use this dataset for joint audio-video generation, since it has been introduced in OmniAudio for the task of video-to-audio generation. 

The dataset comprises over 100k clips ($\sim$288 hours) and ensures a strong alignment between audio and video, which is fundamental for tasks such as audio-visual generation. For our study, we used mono audio representations and focused onto spatial controllability of the visual scene. Since the original clips are 10 seconds long, we divide each into three overlapping 4-second segments (1-second overlap) and we define our own train/validation splits by randomly placing 85\% of samples in the train set and the remaining part in the validation set. 
The dataset is composed of in-the-wild videos taken from YouTube and offers a wide range of acoustic events and diverse audio-visual contexts, providing a solid base for exploring controllable spatial audio-visual generation.

To ensure data consistency within our pipeline, all 360° videos were standardized to the 2:1 aspect ratio required by the SalViT360 saliency model. While we maintained the highest available resolution for each video, any that did not natively conform to the 2:1 format were stretched to meet this requirement.

\subsection{Training Details}

Audio and video U-Nets as well as CMC-PE modules for temporal synchronization are initialized using the weights of our baseline \cite{ishii2024simple} publicy available in the corrisponding official repository. 
In order to keep the training as lightweight and simple as possible but also efficient, we only update the weights of the fundamental controllable generation module introduced by us for spatial and semantic informations. The weights are zero initialized and then optimized using AdamW with a mean squared error (MSE) loss and a learning rate set to 1e-5. 

Regarding the target video samples, we use a frame rate of 8 \textit{fps}, resulting in 32 frames for 4-seconds long videos with 256x256 height and width dimensions; while for audio samples we use mono audio at a sample rate of 16 kHz.
We trained the model for 500 epochs on 4 A100 GPUs.

\subsection{Objective Evaluation}

To quantitatively assess the quality and controllability of the generated audio-visual content, we employ a set of complementary evaluation objective metrics.

We introduce $S_{\text{KL}}$ to specifically measure the spatial alignment of content. This metric quantifies the similarity between the visual attention distributions of the generated video and the target viewpoint. The process involves first estimating a saliency map for both the generated and target videos using a dedicated model \cite{aydemir2023tempsal}. The Kullback-Leibler (KL) divergence between these maps is then calculated on a frame-by-frame basis. The final $S_{\text{KL}}$ score is the average of these individual frame scores, as defined by the formula:
\begin{equation}\nonumber
    S_{\text{KL}} = \frac{1}{T} \sum_{t=1}^{T} \text{KL}(\text{Sal}(\hat{v}_t) || \text{Sal}(v_t)),
\end{equation}
where $T$ is the total number of frames, $\text{Sal}$ is the saliency map and $\hat{v}$ and $v$ are generated and target samples. 

We also use the Fréchet Audio Distance (FAD) \cite{kilgour2018fr}, which evaluates audio fidelity by comparing the feature distributions of real and generated audio, which we compute using Laion-CLAP embeddings. Similarly, we adopt the Fréchet Video Distance (FVD) \cite{unterthiner2018towards} to compute the Fréchet distance between the distributions of features extracted from real and generated video samples, using I3D video encoder. 
    

\begin{table}[t]
    \centering
    \caption{Objective evaluation}
    \begin{tabular}{l c c c c c} \toprule
        Model               & \textit{$S_{\text{KL}}$} $\downarrow$   & FAD $\downarrow$     & FVD $\downarrow$   \\ 
        \midrule
        Baseline \cite{ishii2024simple}        & 1.9536      & 8.23             & 1959              \\         
        Con360-AV  & \textbf{0.6458}     & \textbf{3.21}             & \textbf{1327}              \\
        \midrule
        Viewpoint comparison  & -     & 4.12             & 1594              \\
        \bottomrule 
    
    \end{tabular}
    \vspace{0.3cm}
    \label{tab:objective}
\end{table}

\subsection{Discussion}

We report the result obtained by our model on the objective metrics used in Table \ref{tab:objective}.

To the best of our knowledge, this is the first research work that integrates 360°-aware spatial information for controllable audiovisual joint generation. For this reason, there are currently no other methods against which directly compare our framework for objective evaluation. To have an idea of how much the proposed method can provide a substantial improvement in the spatial controllability of joint generation, we decided to conduct our evaluation comparing the method against our baseline \cite{ishii2024simple}, from which we use the same temporal conditioning but that does not provide any kind of spatial control. 
We follow the same training routine specified by the authors to train the baseline model on the \textit{Sphere360} dataset for a fair comparison. 

Our model demonstrates robust spatial controllability, a claim quantitatively supported by our results. The significantly low $S_\text{KL}$ values achieved by our method indicate strong spatial alignment between the generated content and the target viewpoint and highlight the effectiveness of the conditioning BASD and saliency maps. This precise spatial guidance, coupled with rich semantic conditioning, also leads to an improvement in overall video quality, as measured by the FVD metric. 
\begin{figure}[!t]
    \centering
    \includegraphics[width=\linewidth]{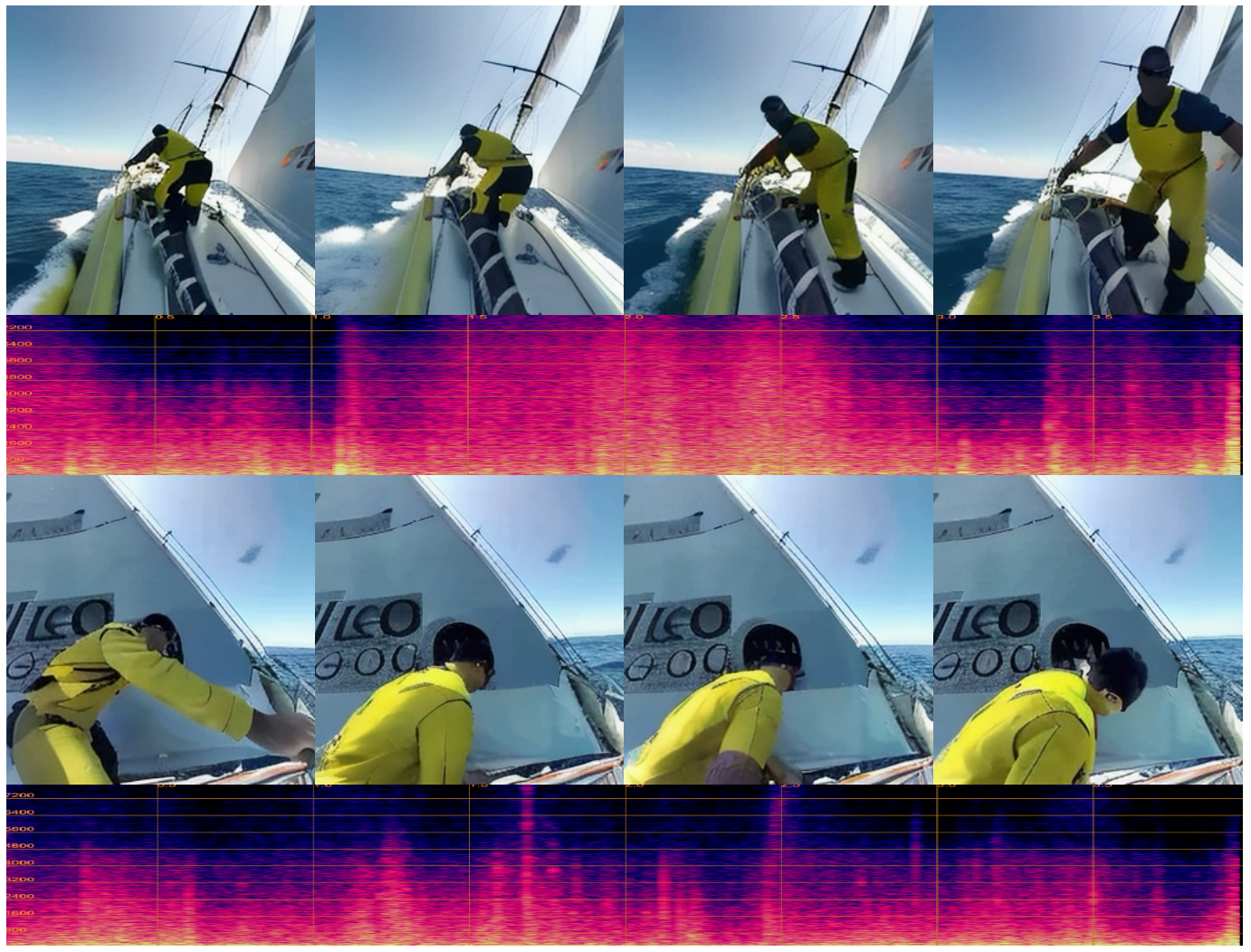}
    \caption{Example of 2 generated viewpoints of the same scene.}
    \label{fig:condmod}
\end{figure}
Furthermore, our parallel joint architecture allows control signals to be injected into the audio and video branches simultaneously. Consequently, the spatial and semantic controls also positively influence the audio generation, resulting in more detailed and contextually aligned audio.
This is evidenced by the improved FAD audio quality metric compared to the baseline. 

We also conduct an additional study to assess how well audio and video from different viewpoints of the same scene are semantically aligned. To do so, we calculate the FAD and FVD quality metrics between samples generated from the same saliency maps and textual description but with different BASD maps. The results, also reported in Table~\ref{tab:objective}, show that the model is able to consistently take off-screen spatial information into account during generation, producing semantically aligned results for different viewpoints of the same scene both in terms of audio and video.
An example of the generated results is reported in Fig.~\ref{fig:condmod}.

%
\section{Conclusion}
\label{sec:conclusion}
We presented Con360-AV, a novel framework for controllable audio-visual generation that leverages panoramic saliency maps, bounding-box-aware signed distance maps, and global captions to generate viewpoint-specific content consistent with the broader 360° spatial context. Our approach introduces explicit spatial controllability, enabling off-screen events to influence the audio-visual scene, which is essential for immersive media applications. Experiments on \textit{Sphere360} have demonstrated that integrating these complementary conditioning signals significantly improves spatial awareness and coherence. In future work, we aim to 
extend our framework to multichannel spatial audio, including Ambisonics, to support fully immersive 3D audio-visual generation.
\section{Acknowledgments}
\label{sec:acknowledgments}
The first author is grateful to Prof. Wenwu Wang for hosting his visit to the University of Surrey, UK, during which this research was conducted.

This work was supported by the European Union under the Italian National Recovery and Resilience Plan (NRRP) of NextGenerationEU, partnership on “National Centre for HPC, Big Data and Quantum Computing” (CN00000013 - Spoke 6: Multiscale Modelling \& Engineering Applications).

\bibliographystyle{IEEEbib}
\bibliography{references}

\begin{thebibliography}{10}

\bibitem{rombach2022high}
R.~Rombach, A.~Blattmann, D.~Lorenz, P.~Esser, and B.~Ommer,
\newblock ``High-resolution image synthesis with latent diffusion models,''
\newblock in {\em Proceedings of the IEEE/CVF conference on computer vision and pattern recognition}, 2022, pp. 10684--10695.

\bibitem{guo2023animatediff}
Y.~Guo, C.~Yang, A.~Rao, Y.~Wang, Y.~Qiao, D.~Lin, and B.~Dai,
\newblock ``Animatediff: Animate your personalized text-to-image diffusion models without specific tuning,''
\newblock {\em arXiv preprint arXiv:2307.04725}, 2023.

\bibitem{liu2023audioldm}
H.~Liu, Z.~Chen, Y.~Yuan, X.~Mei, X.~Liu, D.~P. Mandic, W.~Wang, and M.~D. Plumbley,
\newblock ``Audioldm: Text-to-audio generation with latent diffusion models,''
\newblock in {\em International Conference on Machine Learning}. PMLR, 2023, pp. 21450--21474.

\bibitem{comunita2024syncfusion}
M.~Comunit{\`a}, R.~F. Gramaccioni, E.~Postolache, E.~Rodol{\`a}, D.~Comminiello, and J.~D. Reiss,
\newblock ``Syncfusion: Multimodal onset-synchronized video-to-audio foley synthesis,''
\newblock in {\em ICASSP 2024-2024 IEEE International Conference on Acoustics, Speech and Signal Processing (ICASSP)}. IEEE, 2024, pp. 936--940.

\bibitem{gramaccioni2024folai}
R.~F. Gramaccioni, C.~Marinoni, E.~Postolache, M.~Comunita, L.~Cosmo, J.~D. Reiss, and D.~Comminiello,
\newblock ``Folai: Synchronized foley sound generation with semantic and temporal alignment,''
\newblock {\em arXiv preprint arXiv:2412.15023}, 2024.

\bibitem{GramaccioniIJCNN2025}
R.~F. Gramaccioni, C.~Marinoni, E.~Grassucci, G.~Cicchetti, A.~Uncini, and D.~Comminiello,
\newblock ``Foley{GRAM}: {V}ideo-to-audio generation with {GRAM}-aligned multimodal encoders,''
\newblock in {\em International Joint Conference on Neural Networks (IJCNN)}, Rome, Italy, June 2025.

\bibitem{MarinoniIJCNN2025}
C.~Marinoni, R.~F. Gramaccioni, K.~Shimada, T.~Shibuya, Y.~Mitsufuji, and D.~Comminiello,
\newblock ``Stereo{S}ync: {S}patially-aware stereo audio generation from video,''
\newblock in {\em International Joint Conference on Neural Networks (IJCNN)}, Rome, Italy, June 2025.

\bibitem{liu2024syncflow}
H.~Liu, G.~L. Lan, X.~Mei, Z.~Ni, A.~Kumar, V.~Nagaraja, W.~Wang, M.~D. Plumbley, Y.~Shi, and V.~Chandra,
\newblock ``Syncflow: Toward temporally aligned joint audio-video generation from text,''
\newblock {\em arXiv preprint arXiv:2412.15220}, 2024.

\bibitem{liu2025javisdit}
K.~Liu, W.~Li, L.~Chen, S.~Wu, Y.~Zheng, J.~Ji, F.~Zhou, R.~Jiang, J.~Luo, H.~Fei, and T.-S. Chua,
\newblock ``Javisdit: Joint audio-video diffusion transformer with hierarchical spatio-temporal prior synchronization,''
\newblock in {\em arxiv}, 2025.

\bibitem{ishii2024simple}
M.~Ishii, A.~Hayakawa, T.~Shibuya, and Y.~Mitsufuji,
\newblock ``A simple but strong baseline for sounding video generation: Effective adaptation of audio and video diffusion models for joint generation,''
\newblock {\em arXiv preprint arXiv:2409.17550}, 2024.

\bibitem{laionclap2023}
Y.~Wu*, K.~Chen*, T.~Zhang*, Y.~Hui*, T.~Berg-Kirkpatrick, and S.~Dubnov,
\newblock ``Large-scale contrastive language-audio pretraining with feature fusion and keyword-to-caption augmentation,''
\newblock in {\em IEEE International Conference on Acoustics, Speech and Signal Processing, ICASSP}, 2023.

\bibitem{cicchetti2024gramian}
G.~Cicchetti, E.~Grassucci, L.~Sigillo, and D.~Comminiello,
\newblock ``Gramian multimodal representation learning and alignment,''
\newblock in {\em International Conference on Learning Representations (ICLR)}, 2025.

\bibitem{ruan2023mm}
L.~Ruan, Y.~Ma, H.~Yang, H.~He, B.~Liu, J.~Fu, N.~J. Yuan, Q.~Jin, and B.~Guo,
\newblock ``Mm-diffusion: Learning multi-modal diffusion models for joint audio and video generation,''
\newblock in {\em Proceedings of the IEEE/CVF Conference on Computer Vision and Pattern Recognition}, 2023, pp. 10219--10228.

\bibitem{kim2024versatile}
G.-w. Kim, A.~Martinez, Y.-C. Su, B.~Jou, J.~Lezama, A.~Gupta, L.~Yu, L.~Jiang, A.~Jansen, J.~C. Walker, et~al.,
\newblock ``A versatile diffusion transformer with mixture of noise levels for audiovisual generation,''
\newblock {\em arXiv preprint arXiv:2402.19385}, 2024.

\bibitem{wang2024360dvd}
Q.~Wang, W.~Li, C.~Mou, X.~Cheng, and J.~Zhang,
\newblock ``360dvd: Controllable panorama video generation with 360-degree video diffusion model,''
\newblock in {\em Proceedings of the IEEE/CVF Conference on Computer Vision and Pattern Recognition}, 2024, pp. 6913--6923.

\bibitem{mao2025omniaudio}
Y.~Mao, H.~Liu, J.-B. Liu, X.~Liu, M.~D. Plumbley, and W.~Wang,
\newblock ``Omniaudio: Generating spatial audio from 360-degree video,''
\newblock {\em arXiv preprint arXiv:2404.14906}, 2024.

\bibitem{zhang2024gazefusion}
Y.~Zhang, N.~Wu, C.~Z. Lin, G.~Wetzstein, and Q.~Sun,
\newblock ``Gazefusion: Saliency-guided image generation,''
\newblock {\em arXiv preprint arXiv:2407.04191}, 2024.

\bibitem{simoni2025bounding}
A.~Simoni and F.~Pelosin,
\newblock ``Bounding box-guided diffusion for synthesizing industrial images and segmentation maps,''
\newblock {\em arXiv preprint arXiv:2405.03623}, 2024.

\bibitem{rai2021salvit360}
A.~Rai, C.~Cheng, and Z.~Duan,
\newblock ``Salvit360: A dataset for salient instance segmentation in 360-degree videos,''
\newblock in {\em 2021 IEEE International Conference on Image Processing (ICIP)}. IEEE, 2021, pp. 2843--2847.

\bibitem{li2023blip}
J.~Li, D.~Li, S.~Savarese, and S.~Hoi,
\newblock ``Blip-2: Bootstrapping language-image pre-training with frozen image encoders and large language models,''
\newblock in {\em International Conference on Machine Learning}. PMLR, 2023, pp. 19730--19748.

\bibitem{grattafiori2024llama}
A.~Grattafiori, A.~Dubey, A.~Jauhri, A.~Pandey, A.~Kadian, A.~Al-Dahle, A.~Letman, A.~Mathur, A.~Schelten, A.~Vaughan, et~al.,
\newblock ``The llama 3 herd of models,''
\newblock {\em arXiv preprint arXiv:2407.21783}, 2024.

\bibitem{perez2018film}
E.~Perez, F.~Strub, H.~de~Vries, V.~Dumoulin, and A.~C. Courville,
\newblock ``Film: Visual reasoning with a general conditioning layer,''
\newblock in {\em AAAI}, 2018.

\bibitem{aydemir2023tempsal}
B.~Aydemir, L.~Hoffstetter, T.~Zhang, M.~Salzmann, and S.~S{"u}sstrunk,
\newblock ``Tempsal - uncovering temporal information for deep saliency prediction,''
\newblock in {\em Proceedings of the IEEE/CVF Conference on Computer Vision and Pattern Recognition (CVPR)}, 2023.

\bibitem{kilgour2018fr}
K.~Kilgour, M.~Zuluaga, D.~Roblek, and M.~Sharifi,
\newblock ``Fréchet audio distance: A metric for evaluating music enhancement algorithms,''
\newblock {\em arXiv preprint arXiv:1812.08466}, 2018.

\bibitem{unterthiner2018towards}
T.~Unterthiner, S.~Van~Steenkiste, K.~Kurach, R.~Marinier, M.~Michalski, and S.~Gelly,
\newblock ``Towards accurate generative models of video: A new metric \& challenges,''
\newblock {\em arXiv preprint arXiv:1812.01717}, 2018.

\end{thebibliography}

\end{document}